\documentclass[12pt]{article}
\usepackage{amsmath}
\usepackage{amssymb}
\usepackage{graphicx,bbm,mathrsfs}
\usepackage{color}

\newcommand{\lsim}{\raisebox{-0.13cm}{~\shortstack{$<$ \\[-0.07cm] $\sim$}}~} 
\newcommand{\gsim}{\raisebox{-0.13cm}{~\shortstack{$>$ \\[-0.07cm] $\sim$}}~} 
\newcommand{\beq}{\begin{eqnarray}} 
\newcommand{\eeq}{\end{eqnarray}} 
\newcommand{\tb}{\tan\beta} 

\begin{document}

\begin{flushright}LPT-Orsay--15--99 \end{flushright} 

\begin{center}

{\large\bf 
Scenarii for interpretations of the LHC diphoton excess:}

\vspace*{2mm}

{\large\bf 
two Higgs doublets and vector--like quarks and leptons}

\vspace*{4mm}

{\sc Andrei~Angelescu, Abdelhak~Djouadi} and {\sc  Gr\'egory Moreau} 

\vspace{4mm}

{\small 
Laboratoire de Physique Th\'eorique,  CNRS and Universit\'e Paris-Sud \\  
B\^at. 210, F--91405 Orsay Cedex, France \\
}

\end{center}

\begin{abstract}

An evidence for a diphoton resonance at a mass of 750 GeV has been observed in the data collected at the LHC run at a center of mass energy of 13 TeV. We explore several interpretations of this signal in terms of Higgs--like resonances in a two--Higgs doublet model and its supersymmetric incarnation, in which the heavier CP--even and CP--odd states  present in the model are produced in gluon fusion and decay into two photons through top quark loops. We show that one cannot accommodate the observed signal in the minimal versions of these models and that an additional particle content is necessary. We then consider the possibility  that vector--like quarks or leptons strongly enhance the heavy Higgs couplings to photons and eventually gluons, without altering those of the already observed 125 GeV state. 
 
\end{abstract}

\subsection*{1. A diphoton resonance at 750 GeV?}

It has been reported that the approximately 4 fb$^{-1}$ of data, delivered in the latest LHC 
run with a center of mass energy of 13 TeV, indicate the presence of a resonance 
that decays into two photons,  with a mass of about 750 GeV and a width of $\sim 50$ GeV
\cite{annonce}. The local significance of this signal is only at the 3$\sigma$ level in the case of the ATLAS collaboration and slightly less for the CMS collaboration. Hence, it is likely that this excess of data is simply yet another  statistical fluctuation which will be washed away with more data. Nevertheless, in the absence of any firm sign of the long awaited 
new physics beyond the SM, it is interesting to contemplate that the effect is indeed
due to a new resonance. 

Let us briefly sketch the various possibilities for such a resonance and consider its
spin--parity quantum numbers.  The observation of the $\gamma\gamma$ signal rules out the 
option that it comes from the decay of a  spin--1 particle by virtue of the Landau--Yang theorem \cite{Landau-Yang}. This leaves the spin 0 and spin $\geq 2$ possibilities. A graviton--like spin--2 is extremely unlikely since it has universal couplings and it should have also decayed into other states such as $WW, ZZ$, dileptons and dijets which have not been observed up to very high masses \cite{Latest-13TeV}.  The most likely possibility for the resonance particle is thus to have spin--0. 

Furthermore, the resonance should be Higgs--like and couple only very weakly to light quarks as,  if produced in $q\bar q$ annihilation, it should have already been observed at the first LHC run with $\sqrt s=8$ TeV and 20 fb$^{-1}$ data. Indeed, for a 750 GeV resonance, just considering naively the $q\bar q$ parton luminosities for a given c.m. energy, there should be an increase of only a factor of 2.5 for the production rate when moving from 8 to 13 TeV c.m. energy \cite{LHC-XS}.  Instead, if the resonance is produced in $gg$ fusion, the jump 
in cross section would be a factor of 4.5 so that the collected data sets at
$\sqrt s=8$ and 13 TeV would be equivalent. Note that there was an ATLAS search for a two-photon scalar resonance performed at $\sqrt s=8$ TeV \cite{ATLAS-earlier}  but it did not extend beyond the scale of 600 GeV.   CMS searched also for diphoton resonances \cite{CMS-earlier} and observed a slight excess of about $2\sigma$ at a mass of 750 GeV. The present diphoton excess is, thus, not a complete surprise.

In this note, we explore the possibility that the diphoton events originate from the decays  
of the heavy neutral CP--even and CP--odd Higgs particles that are present in two Higgs doublet models \cite{2HDM,HHG} and, in particular, in their minimal supersymmetric incarnation,  the MSSM \cite{HHG,Review2}. We show that to achieve such a strong diphoton signal, additional particles should contribute to the loop induced production and decay processes  and we investigate some scenarii with vector--like quarks and leptons that occur in many extensions of the Standard Model (SM) \cite{VLQs}. We show that indeed, vector--like leptons can account for the observed signal without altering the properties of the lightest $h$ boson.  

 
\subsection*{2. The diphoton rate in 2HDMs and the MSSM}

A straightforward extension of the SM that involves additional Higgs states is a (2HDM) model
with two--Higgs doublets $\Phi_u$ and $\Phi_d$ \cite{2HDM,HHG}  that leads to the presence of four additional physical Higgs bosons besides the 125 GeV state already observed and that we will denote by $h$: a heavier CP--even $H$ state, a CP--odd $A$ state and two charged $H^\pm$ bosons. The model is described by the four Higgs masses $M_h, M_A, M_H$ and $M_{H^\pm}$, and two angles: the angle $\beta$ given by the ratio of the  vacuum expectation values of the two scalar fields and  the mixing angle $\alpha$ that diagonalises the two CP--even Higgs states
\beq 
\tan\beta=v_u/v_d \ \  {\rm and}  \ \ \Phi_u^0 = \sin\alpha\ H + \cos\alpha\ h  \, , \,  
\Phi_d^0=\cos\alpha\   H - \sin\alpha\  h 
\label{eq:alpha}
\eeq
To make in a natural way that the lighter $h$ state is SM--like as experimentally observed \cite{HiggsCombo}, one can nevertheless invoke the alignment limit \cite{alignment} in which the two angles are related by $\beta\!-\! \alpha \! =\! \frac{\pi}{2}$. The $h$ couplings to up and down type fermions and gauge bosons, normalised to the SM values, are then SM--like,  $g_{hVV}\!  \approx \! g_{huu} \! \approx \! g_{hdd} \! \approx \! 1$, while the couplings of the CP--even $H$ and charged $H^\pm$ states  reduce to those of the pseudoscalar $A$. The $H$ coupling to the $V=W,Z$ bosons tends then to zero, $g_{HVV}=\cos(\beta-\alpha)\to 0$,  as for the CP--odd Higgs state which, because of CP--invariance,  has no $VV$ couplings, $g_{AVV} =0$. Depending on whether the up and down--type fermion masses are generated by only one or both Higgs fields, the normalised $\Phi=H,A$ couplings to fermions in the  alignment limit are given by $g_{\Phi bb} = - \tan \beta$ for type II and $g_{\Phi bb} = 1/\tan \beta$ for type I models while one has $g_{\Phi tt} = 1/\tan \beta$ in both scenarios.  The couplings of $\tau$--leptons follow that of $b$-quarks.

The minimal supersymmetric extension of the SM (MSSM) is a type II 2HDM but it has the interesting feature that supersymmetry imposes strong constraints on the parameters and only two of them, e.g. $\tan \beta$ and $M_A$, are independent. This is true not only at tree level but approximately also at higher orders if the constraint $M_h=125$ GeV, which fixes the important radiative corrections to the Higgs sector \cite{Review2}, is enforced; this is the so--called $h$MSSM discussed recently \cite{hMSSM}. Because the LHC Higgs data indicate that $h$ state is SM--like \cite{HiggsCombo} and that the pseudoscalar Higgs boson should be rather heavy \cite{LHC-tautau}, one is also in the so--called decoupling limit in which one has $\cos(\beta-\alpha) \approx 0$ and $M_H \approx M_{H^\pm} \approx M_A \gg M_Z$. This simplifies considerably the phenomenology of the model. 

In this section, we will consider both the  MSSM in the decoupling limit and the 2HDMs of type
I and II in the alignment limit. In the latter case, we will assume in addition that the CP--even $H$ and CP--odd $A$ states are approximately degenerate in mass, $M_A \approx M_H$,  as is the case in the MSSM.  Hence, the 750 GeV diphoton resonance will consist of both the $H$ and $A$ bosons. As motivated below, we will specialize in the low $\tb$ region,  $\tb \lsim 1$, which allows for strong Yukawa couplings of the top quark. The possibility of extremely large Yukawa for bottom quarks requires large $\tb$ values (for type II 2HDMs), $\tb \gsim 30$ and even 50, which for $M_\Phi \approx 750$ GeV, are excluded by the $\Phi \to \tau^+ \tau^-$ searches performed at the run I of the LHC \cite{LHC-tautau}.  Hence, only the top quark Yukawa coupling will be kept in our following discussion and all other Yukawa couplings will be considered to be  negligible and ignored. 

\begin{figure}[!h]
\vspace*{-2.3cm}
\begin{center}
\hspace*{-2cm}
\resizebox{1.0\textwidth}{!}{\includegraphics{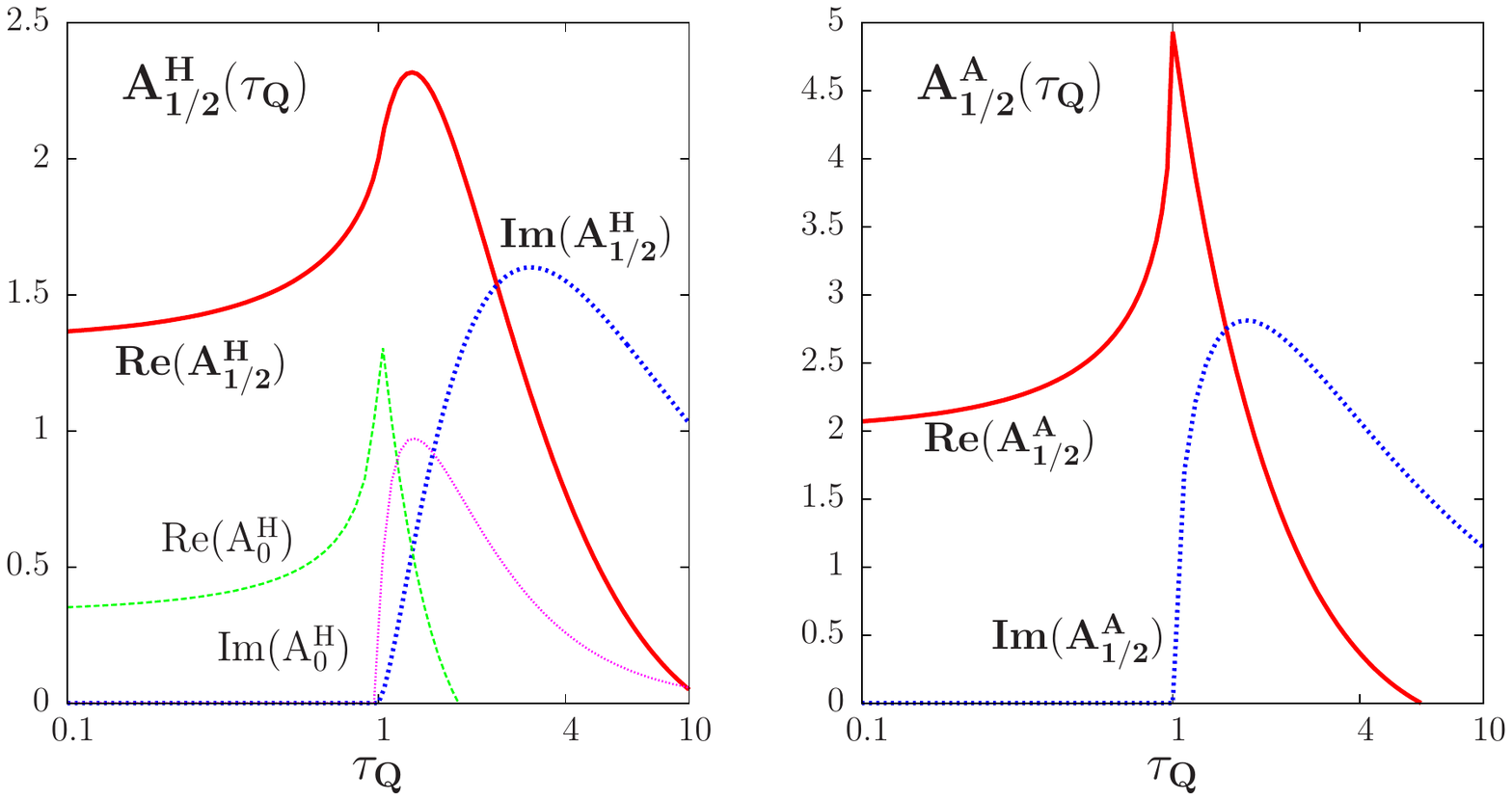}}
\end{center}
\vspace*{-13.5cm}
\caption{The form factors $A^\Phi_{1/2}$ of the Higgs--$gg$ and Higgs--$\gamma\gamma$ 
fermion loops in the case of the CP--even (left) and CP--odd (right) Higgs 
particles as a function of $\tau_f=M_\Phi^2/4m_f^2$. The smaller form factor for
spin--0 particle exchange in the $H$ case is shown for comparison.} 
\label{Fig:Afactor}
\vspace*{-2mm}
\end{figure}

Let us now discuss the decays of the two Higgs resonances. In the configuration that we have chosen, with large Higgs masses and low $\tb$ values, the only relevant tree level decay of the $\Phi=H,A$ bosons will be into top quark pairs with a partial width
\cite{Review1}
\begin{eqnarray}
\Gamma(\Phi \to t \bar{t} ) =  \frac{3 G_\mu m_t^2}{4\sqrt{2} \pi}
\, g_{\Phi tt}^2 \, M_{\Phi} \, \beta^{p_\Phi}_t
\end{eqnarray}
where $\beta_t=(1-4m_t^2/M_{\Phi}^2)^{1/2}$ is the quark velocity and  $p_\Phi =3\,(1)$ 
for the CP--even (odd) Higgs boson. In principle, the decays of the  two resonances into two photons also proceeds through the top--quark loop only (for the CP--even $H$ state, we ignore
the $W$--loop contribution and  there are also small charged Higgs contributions to be discussed shortly), but we will allow for additional contributions of new fermions that we will explicit later. The partial decay widths are given by \cite{HHG,Review1}
\begin{eqnarray}
\Gamma(\Phi  \to \gamma\gamma) =  \frac{G_\mu\alpha^2 M_\Phi^3}
{128\sqrt{2}\pi^3} \bigg| \frac43  g_{\Phi tt} A_{1/2}^\Phi  (\tau_t) +  
{\cal A}^\Phi_{\rm new} 
\bigg|^2
\end{eqnarray}
The form factors $A_{1/2}^\Phi (\tau_f)$ which depend on the variable $\tau_f=M_\Phi^2/
4m_f^2$ are the only place in which the $H$ and $A$ states behave differently. They are shown 
in Fig.~\ref{Fig:Afactor}.  While the amplitudes are real for $M_\Phi \le 2m_f$, they develop an imaginary part above the kinematical threshold. At low Higgs masses compared 
to the internal fermion mass, $M_\Phi \ll 2 m_f$, the amplitudes for a scalar and a pseudoscalar\footnote{Note that when including the QCD corrections to the quark loops,
there is a Sommerfeld enhancement of the amplitudes near threshold which is significant 
in the case of the pseudoscalar Higgs boson. Outside this threshold the QCD corrections are very small; see Ref.~\cite{SDGZ}.}  states reach constant but different values $A_{1/2}^H (\tau_f) \to 4/3$ and  $A_{1/2}^A (\tau_f) \to 2$. The maximal values of the amplitudes occur at the kinematical threshold $M_\Phi = 2 m_f $,  where one has  ${\rm Re} (A^H_{1/2}) \sim 2$ and  ${\rm Re}(A^A_{1/2}) \sim 5$ for the real parts. For a resonance with a mass $M_\Phi \approx 700$ GeV,  one has $\tau_t \approx 4$ for the top quark and this leads to values for the form factors of  ${\rm Re} (A^H_{1/2}) \approx \frac34 ,  {\rm Im}(A_{1/2}^H) \approx \frac32$ in the CP--even case and  ${\rm Re} (A^A_{1/2}) \approx \frac13  ,  {\rm  Im}(A_{1/2}^A) \approx 2$ in the CP--odd case. 

Assuming that there are no new physics contributions to the $\Phi \gamma \gamma$ loop
besides that of the top quark, the branching ratio of the decay $\Phi \to \gamma\gamma$ 
is simply given by 
\beq 
{\rm BR}(\Phi \to \gamma\gamma) \simeq \frac{ \Gamma (\Phi \to \gamma\gamma) }
{ \Gamma (\Phi \to t\bar t)}=  \frac{\alpha^2}{54 \pi^2} \frac{M_\Phi^2}{m_t^2} \frac{
|A_{1/2}^\Phi|^2}
{\beta_t^{p_\Phi}} \approx 10^{-7} \; \frac{M_\Phi^2}{m_t^2} \; \frac{1}{\beta_t^{p_\Phi}}
\eeq
and does not depend on $\tan\beta$. For $M_\Phi \approx 750$ GeV, one obtains fractions 
BR$(A \to \gamma\gamma) \approx 7 \times 10^{-6}$ and BR$(H \to \gamma\gamma) \approx 6 
\times 10^{-6}$ and total decay widths, with $\Gamma_{\rm tot}^\Phi \sim \Gamma (\Phi \to 
t\bar t)$, of $\Gamma_{\rm tot}^H \approx 32\;{\rm GeV} /\tan^2\beta$ and $\Gamma_{\rm tot}^A \approx 35\;{\rm GeV} /\tan^2\beta$ \cite{hdecay}. To arrive at a total width of 
$\approx 50$ GeV  as experimentally observed, one thus needs $\tan \beta \! \approx \! 1$.
Hence, besides requiring the equality\footnote{Note that if we assume the $h$MSSM with $M_A \approx 750$ GeV, this leads $M_H \approx 765$ GeV for $\tan\beta \approx 1$ \cite{hMSSM}; in a 2HDM, the $H/A$ masses can also be different. A significant $M_H-M_A$ difference will make the observed resonance wider, which is not our interest here, so we assume equal masses also in 2HDMs.} $M_H \approx M_A$, on should not allow  for new $H/A$ decay channels in order not to increase this total width. In fact, even if the total width issue could be ignored,  $\tb$ values much smaller than unity, say $\tb \! \lsim \! 1/3$, should be avoided in order to keep
a perturbative top quark Yukawa coupling. We will thus stick to $\tb \approx 1$ in our analysis. 

In  a similar way, the cross section for $\Phi$ production in the dominant gluon--gluon fusion process is proportional to the Higgs decay width into two gluons which is given by
\cite{HHG,Review1} 
\begin{eqnarray}
\sigma  (gg\to \Phi) \propto \Gamma(\Phi  \to gg) =  \frac{G_\mu\alpha_s^2 M_\Phi^3}
{64\sqrt{2}\pi^3} \bigg|   g_{\Phi tt} A_{1/2}^\Phi  (\tau_t) 
+ {\cal A}^\Phi_{\rm new} 
\bigg|^2
\end{eqnarray}
First, one notices that the production cross section at $\sqrt s=13$ TeV for a SM--like Higgs boson of mass  $M_{\rm H}=750$ GeV is $\sigma (H_{\rm SM}) \approx 0.85$ pb \cite{hMSSM} and that in our case, one has $\sigma (H)/\sigma (H_{\rm SM}) = \cot^2\beta$ and $\sigma (A)/\sigma (H_{SM})= \cot^2 \beta \times |A_{1/2}^A/ A_{1/2}^H |^2 \approx 2 \cot^2\beta$ as the form factor is different in the CP--odd case. One obtains then for the cross section times branching fraction when the two channels are added (the numbers are for the $h$MSSM),
\beq 
\sum_\Phi \sigma (gg\to \Phi) \times {\rm BR}( \Phi \to \gamma\gamma) \approx 
1.5 \times 10^{-2} \cot^2\beta  \; {\rm [fb]}
\eeq
to be compared with a cross section  of ${\cal O}(10\, {\rm fb})$ observed by the ATLAS collaboration. 

Thus, for $\tb \approx 1$, we are more than two orders of magnitude  away from 
the diphoton signal and, even if we allow for $\tb \approx 1/3$, we are still more than an 
order of magnitude below. Very large additional contributions are thus needed. 

An important remark is that if the enhancement of the diphoton signal had to come from the production cross section, then we would have had a large rate for $gg \! \to \! \Phi \! \to
\! t\bar t$ that is constrained from the search of resonances decaying into top quark pairs at the 8 TeV LHC. Indeed a 95\% confidence limit of $\sigma(gg\to \Phi) \times {\rm BR}(\Phi \to t\bar t) \lsim 1~{\rm pb}$~\cite{LHC-ttbar} has been set and, since at this energy one has $\sigma(gg\! \to\! H\!+\!A) \approx 0.5 \cot^2\beta$ pb and ${\rm BR}(\Phi \to t\bar t) \approx 1$, $\tb$ cannot take values much smaller than unity. This leads to the important conclusion that the two orders of magnitude enhancement needed to accommodate the observed diphoton resonance in our context should essentially come from the $\Phi \to \gamma\gamma$ decay. 

Although obviously very unlikely, we nevertheless considered the various additional contributions that can affect the $\Phi \gamma\gamma$ and $\Phi gg$ loops in the minimal versions of 2HDMs and the MSSM and checked that such a huge enhancement cannot be obtained (this is clearly also the case for the lightest $h$ boson as it has recently been discussed in Ref.~\cite{golden}). 

A first contribution to the $\Phi \gamma\gamma$ loops which can be considered,  is that of a charged Higgs boson in the case of $H$ (because of CP--invariance there is no $AH^+H^-$ coupling) given by ${\cal A}_{H^\pm}^H= g_{H H^+ H^-} (M_W^2/ M_{H^\pm}^2) \times A_0^H(\tau_{H^\pm})$. The form factor $A_0^H$ is smaller compared to the fermionic one as shown in Fig.~\ref{Fig:Afactor}. In the MSSM, as $M_{H^\pm} \approx M_H$ and $g_{HH^+ H^-} ={\cal O}(1)$, the contribution is negligible. Even in a  general 2HDM, although $g_{HH^+ H^-}$ is not fixed and can be made relatively large, the $H^\pm$ contributions are also very small\footnote{For $M_{H^\pm} \approx 160$ GeV which satisfies the constraints set at the 8 TeV LHC \cite{PDG}, the form factor is small and negative giving a destructive interference with the top contributions.  Instead, the contribution can be increased by sitting close to the $M_{H^\pm}= \frac12 M_\Phi$ threshold so that $A_0^H$ reaches it maximal value, Re$(A^H_{0}) \sim 1.5$ and  Im$(A^H_{0}) \sim 1$ for $\tau \sim 1$. Still these values are too small and the damping factor $M_W^2/M_{H^\pm}^2$ too strong and even for extremely large $g_{HH^+ H^-}$, the 
contributions stay modest.}. 
 
In the case of the MSSM, additional contributions are provided by supersymmetric particles
running in the loops. The contributions of the charginos in $\Phi \to \gamma \gamma$ are in general small if we are above the $M_\Phi \gsim 2m_{\chi^\pm}$ thresholds that are needed to keep the total decay widths of  the resonances small. But also for small chargino masses, BR$(\Phi \to \gamma\gamma)$ cannot be enhanced by more than a few ten percent\footnote{For the choice of SUSY parameters $\tan\beta=1, M_2=-\mu=200$ GeV which leads to $\chi_1^\pm$ with masses close to the experimental bounds $M_{\chi_1^\pm} \approx 100$ GeV \cite{PDG}, and maximally coupled to the $H/A$ states, one makes only a 25\% and 10\% change of the $H\to \gamma \gamma$ and $A\to \gamma \gamma$ branching ratios respectively \cite{hdecay}.}. There also contributions of sleptons and squarks to the CP--even $H \to \gamma \gamma$ decay and squarks to $gg\to H$ production;  the CP--odd $A$ state does not couple to identical sfermions and there is no contribution at lowest order. Here again, the Higgs--sfermion couplings are not proportional to sfermions masses and the contributions, $\mathcal{A}^H_{\tilde f } \propto \sum_{\tilde f_i} g_{H \tilde f_i \tilde f_i}/ m_{ \tilde f_i}^2 \times A^H_0 (\tau_{\tilde f} )$,  are damped by powers of $ m_{ \tilde f_i}^2$ leading to small loop contributions for sufficiently heavy sfermions. This is particularly true in the slepton case where the dominant contribution due to light stau's cannot be enhanced by strong couplings for the low $\tb$ values that we are considering here.  

Finally squarks and particularly relatively light top squarks can make significant contributions to $H\to \gamma \gamma$ and $gg\to H$. In the MSSM, however, for the low $\tb$ values that we are considering,   the stops (which contribute to the radiative corrections that enhance the lighter $h$ boson mass)  should be extremely heavy for $M_h=125$ GeV to be reached. Even if by some means one can accommodate this mass  values with light stop (e.g. by invoking an additional singlet--like Higgs as in the NMSSM or by incorporating some additional particles to increase the radiative corrections to the $h$ mass) it is very difficult to increase $\sigma(gg \! \to \! H) \! \times \! {\rm BR} (H \!\to \! \gamma \gamma)$ significantly\footnote{For instance,  assuming $m_{\tilde t_1} \approx m_{\tilde t_2} \approx \frac12 M_H \approx 350$ GeV in order to maximize the $A_0^H$ amplitude, one obtains  a factor of $\approx 2$ enhancement of $\sigma(gg \! \to \! H) \! \times \!{\rm BR} (H \! \to \!\gamma \gamma)$. Instead, for a trilinear stop coupling $A_t\! \approx  \!2$ TeV that strongly enhances the coupling $g_{H \tilde t_1 \tilde t_1} \propto m_t A_t$, one obtains a more modest change \cite{hdecay}.}. In fact, in general, when the SUSY contributions are large, they are also large in the case of the lightest $h$ boson \cite{golden} which is unacceptable as its couplings have been measured to be SM--like. 

In conclusion, it is very difficult to enhance the production cross section and the $\gamma \gamma$ decay branching ratios of the MSSM $H$ and $A$ bosons to a level close to what is experimentally observed, even if extreme configurations for the superparticle masses and couplings are chosen. Other, more radical, measures are needed and we discuss them now.


\subsection*{3. Introducing vector--like quarks and leptons}

In order to increase significantly the Higgs couplings to gluons and/or photons, one 
could consider the contributions of new heavy fermions to the triangular loops\footnote{One can of course consider also the introduction of scalars, such as doubly charged Higgs bosons from see-saw mechanisms \cite{DCHiggs} for instance, but we will not consider this option here.}. These fermions should have vector--like couplings to the electroweak gauge bosons in order to avoid genera\-ting their masses through the Higgs mechanism only and then cope with the  electroweak precision tests as well as the LHC Higgs data\footnote{The easiest option would have been the introduction of a fourth generation of fermions, which could have increased  both the $gg\to H/A$ cross section and the $H/A \to \gamma\gamma$  decay rates by an order of magnitude each. This is nevertheless ruled out by the observation of the light $h$
state with SM--like couplings \cite{fourth}.}.  Vector--like fermions appear in many extensions of the SM and recent discussions have been given in Ref.~\cite{VLQs}. In our analysis, we will not rely on any specific model (as e.g.~Ref.~\cite{SUSY+VLQ}) and simply adopt an effective approach in which the properties of these fermions are adjusted in order to fit our purpose. 

In addition to the two Higgs doublets for which we still assume the alignment limit and the mass equality $M_H \approx M_A$, we first consider vector--like quarks (VLQs) with the following minimal Lagrangian describing their Yukawa couplings in the interaction basis 
\begin{align} 
- {\cal L}_{\rm VLQ}  &=  \bigg \{ \frac{y_{L}^{b}}{\sqrt{2}} 
\! \left( \! \begin{array}{c} 0  \\   v_d\!+\!\Phi_d^0\!+\!iP_d^0 \end{array} \! \right)\!
\overline{ \left( \! \begin{array}{c} t'  \\   b' \end{array}  \! \right)}_{ \hspace*{-2mm}L\;} \!  b''_{R} 
 +   \frac{y_{L}^{t}}{\sqrt{2}} \! \left( \! \begin{array}{c}
v_u\! + \! \Phi_u^0 \! \pm \! i P_u^0  \\   0
\end{array}  \! \right) \! \overline{  \left(  \begin{array}{c}
t'  \\   b' \end{array}  \! \right)}_{\hspace*{-2mm} L\; } t''_{R}   +  
 \{ {\rm  L \! \leftrightarrow \! R }   \}    \hspace*{-1cm} \notag \\ 
\ \ &+ \ m_1 \overline{ \left( \begin{array}{c} t'  \\ b'  
\end{array}  \right)}_{ \hspace*{-1mm} L\; } \left ( \begin{array}{c} t'  \\ b'   
\end{array} \right)_{ \hspace*{-1mm} R \; }
\ + \ m_2 \ \overline{t''_{L}}  t''_R  \ + \ m_3 \ \overline{b''_{L}} b''_R \bigg \}
\ + \ {\rm h.c.}, 
\label{eq:LagVLQ}
\end{align}
where the ``$+$" sign corresponds to the MSSM case, while the ``$-$" sign corresponds to a type II 2HDM case\footnote{This difference comes from the fact that in the MSSM, because of the holomorphicity of the superpotential, the field $\Phi_u$ with hypercharge $-\frac12$ has to give mass to the up-type quarks, whereas in the type II 2HDM considered here one can use $\Phi_u \equiv \tilde{\Phi}_2 = i \sigma_2 \Phi_2^*$, where $\Phi_2$ has $+\frac12$ hypercharge (as $\Phi_1 = \Phi_d$).}. Such representations of the VLQs make possible the presence of Yukawa couplings invariant under the SM gauge symmetry, hence including also terms mixing the VLQs with SM quarks. However, this mixing would  represent only a higher order correction to the main enhancement effect of interest; we  have therefore omitted for simplicity such Yukawa couplings in the Lagrangian of eq.~(\ref{eq:LagVLQ}). The key point here is that, to fulfill gauge invariance, at least two vector-like multiplets need to be introduced in order to generate direct Yukawa couplings for the VLQ that are not suppressed by SM--VLQ mixing angles. This will allow to have significant VLQ loop contributions. The motivation to have both states of type $t^{\prime(\prime)}$ and $b^{\prime(\prime)}$, coupling respectively to the $\Phi^0_u$ and $\Phi^0_d$ scalar fields, will become clear later. Such a configuration is also motivated by economy since it follows that of the SM. We will further consider six families of the above VLQ multiplets, to obtain the diphoton rates that are compatible with the LHC data.

Similarly to VLQs, one can introduce vector--like leptons (VLL) in the model. The VLLs are subject to weaker direct mass bounds \cite{VLLEP} than the ones of order 1 TeV on the VLQ masses \cite{LHC-VLQS, VLQ-india}. For the particle content, an interesting possibility would be  to introduce several replica of vector--like lepton doublets and singlets
\beq
\left( \!  \begin{array}{c} \ell^{\prime -}  \\  \ell^{=} \end{array}  \! \right)_{ \hspace*{-2mm}L/R\;}  , \   
\ell^{\prime\prime -}_{L/R} \ , \ \ell^{\prime =}_{L/R},  
\label{eq:VLL-ii}
\eeq
which will couple to the 2HDM Higgs states exactly as shown in the Lagrangian of eq.~(\ref{eq:LagVLQ}) with the replacement $ t', t'' \to \ell^{\prime -} , \ell^{\prime\prime -}$ 
and  $b', b'' \to  \ell^{=} , \ell^{\prime =}$. The reason to consider this specific pattern for the ${\rm SU(2)_L}$ doublet, that includes a singly and a doubly charged lepton, 
is that it allows both its components to contribute to the diphoton triangular loop\footnote{ 
This would not have been the case of a vector--like lepton doublet with a neutrino and 
a singlet charged lepton. However, in the same spirit of Higgs portal models for dark matter \cite{DM}, one could introduce a doublet $(\nu',\ell')_{L/R}^t$ with a gauge singlet $\nu''_{L/R}$ protected from decays by a parity forbidding Yukawa coupling between these extra leptons and the SM ones (and in turn their mixings). In the present 2HDM framework, the coupling of $\nu^{\prime(\prime)}_{L/R}$ to the neutral Higgs states, through its initial coupling to $\Phi_u^0$, $P^0_u$, would open up the possibility of an  invisible decay width for the two heavy Higgs bosons as well.}.  
 
Another possibility would be for instance to introduce instead the representations $(\ell^{=} , \ell^{\equiv})^t_{L/R} , \ell^{\prime =}_{L/R} ,  \ell^{\prime \equiv}_{L/R}$, with couplings as in eq.~(\ref{eq:LagVLQ}) with $ t', t'' \to \ell^{=} , \ell^{\prime =}$ and  $b', b'' \to  \ell^{\equiv} , \ell^{\prime \equiv}$.  The motivation would be that these VLLs with higher electric charges would lead to a much stronger enhancement of  Higgs--diphoton vertices.

Based on the particle content of eqs.~(\ref{eq:LagVLQ})--(\ref{eq:VLL-ii}) and recalling the Higgs eigenstate compositions of eq.~(\ref{eq:alpha}) that involves the mixing angle $\alpha$  which in the alignment or decoupling limits is given by $\alpha=\beta-\frac{\pi}{2}$, one can express the new fermion contributions to the loop induced $\Phi=H,A$ couplings to photons and gluons in the following form, 
\beq
{\cal A}^\Phi_{\rm VLF} (gg) & \propto & {\cal A}^\Phi_{\rm top} ( gg) +  N^{\rm VLQ}_f  
\left ( \cot\beta \sum^2_{i=1}\frac{v g^{
\Phi}_{t_{ii}}}{m_{t_i}}A^\Phi_{1/2} (\tau_{t_i}) \pm
\tan\beta\sum^2_{i=1}\frac{v g^{
\Phi}_{b_{ii}}}{m_{b_i}}A^\Phi_{1/2} (\tau_{b_i}) \right ),~~~
\label{eq:agg} \\ 
{\cal A}^\Phi_{\rm VLF} (\gamma\gamma) & \propto & {\cal A}^\Phi_{\rm top} (\gamma \gamma) +  
{\cal A}^\Phi_{W} (\gamma \gamma)  
+\! N^{\rm VLF}_f \Bigg( \cot \beta \ \sum_u N^u_c Q_u^2 \sum^2_{i=1} \frac{v g^{
\Phi}_{u_{ii}}}
{m_{u_i}}A^\Phi_{1/2} (\tau_{u_i}) \nonumber\\ &\pm&  
\tan\beta \ \sum_d N^d_c Q_d^2 \sum^2_{i=1}\frac{v g^{
\Phi}_{d_{ii}}}{m_{d_i}}A^\Phi_{1/2} (\tau_{d_i}) \Bigg),
\label{eq:app}
\eeq
where we have slightly changed the notation. The ``$+$" sign corresponds to $\Phi = A$ in the type II 2HDM, while the ``$-$" sign corresponds to all the other cases, namely $\Phi = H$ in the type II 2HDM and $\Phi=H,A$ in the MSSM. In order to describe generically either VLQ or VLL, we introduce sums over $u$ ($d$) to account for fermions with different electric charges coupling to $\Phi_u^0 $ ($ \Phi_d^0 $), while $i=1,2$ indicate the heavy lepton mass eigenstates. For up--type and down--type vector--like leptons, the mass eigenvalues $m_{u_i,d_i}$ are obtained from bidiagonalizing the mass matrices
\begin{equation}
\mathcal{M}_u = \left ( \begin{array}{cc}
m_1 & \frac{1}{\sqrt{2}} v y_L^u \sin\beta \\ \frac{1}{\sqrt{2}}v y_R^u \sin\beta & m_2
\end{array} \right ), \quad 
\mathcal{M}_d = \left ( \begin{array}{cc}
m_1 & \frac{1}{\sqrt{2}} v y_L^d \cos\beta  \\ \frac{1}{\sqrt{2}} v y_R^d \cos\beta & m_2
\end{array} \right ),
\label{Eq:mass_matrix}
\end{equation}
in the $(\ell^{\prime -}, \ell^{\prime\prime -})_{L/R}$ and $(\ell^{=},\ell^{\prime=})_{L/R}$ bases respectively. The $g^{\Phi}_{u_{ii},d_{ii}}$ denote, up to possible factors of $\tan\beta$ and $\cot\beta$, the diagonal elements of the mass basis Yukawa coupling matrix of $\Phi=H,A$ to the VL states. The aforementioned coupling matrices are obtained from the biunitary transformations of 
\begin{equation}
y_{u} = \frac{\sin\beta}{\sqrt{2}} \left( \begin{array}{cc}
0 & y_L^u \\ \pm y_R^u & 0
\end{array} \right), \quad 
y_{d} = \frac{\cos\beta}{\sqrt{2}} \left( \begin{array}{cc}
0 & y_L^d \\ \pm y_R^d & 0
\end{array} \right),
\label{eq:yukawas}
\end{equation}
with ``$+$" for the $h, H$ states and ``$-$" for the $A$ state\footnote{We thank Xiao-Fang Han for pointing out to us an inconsistency in the couplings of the pseudoscalar $A$ boson in an earlier version of the paper.}. As usual, $N_c$ and $Q_f$ are the color and electric charges whereas $N_f^{\rm VLF}$ stands for the number of vector--like fermion families (taken here, for simplicity, decoupled one from another).  The top quark contributions to the loops are simply given by ${\cal A}^{H,A}_{\rm top} (\gamma \gamma) =  \frac43 \cot\beta A_{1/2}^\Phi (\tau_t)$ and ${\cal A}^{H,A}_{\rm top} (gg) = \cot\beta A_{1/2}^\Phi (\tau_t)$. Also, we have introduced a $W$ contribution to the $\gamma\gamma$ amplitude,  which, in the decoupling limit, is zero in the case of the $H/A$ states, but not of the lighter $h$ state.  

In the narrow width approximation, fully justified for resonances with total decay width of about $5\%$ of their masses, the production rates $\sigma(gg\! \to \! \Phi \! \to \! \gamma \gamma)$ are simply proportional to the product of the two amplitudes squared, $\sigma \! 
\times \! {\rm BR} \! \propto \! |{\cal A}^\Phi (\gamma\gamma)|^2 \! \times \!  |{\cal A}^\Phi (gg)|^2$. 
 
A very important requirement for the vector--like content contributing to the $\Phi gg$ and $\Phi \gamma\gamma$ couplings is to not alter significantly the absolute values of the loop induced couplings of the lightest $h$ state with a mass of 125 GeV, which has already been observed to be approximately SM--like~\cite{HiggsCombo}. Indeed, the amplitudes for $h$ are almost exactly the same as those of $H$ given by eqs.~(\ref{eq:agg})--(\ref{eq:app}) with now the non--zero $W$ loop contribution included in the diphoton case.  An important difference though is that one has to make the replacement $\pm\tan\beta \to 1$ and $\cot\beta \to 1$. This will turn out to be essential for reproducing the measured value for the effective $h\gamma\gamma$ ($hgg$) coupling.

For the values of the parameter $\beta$ that we are considering  and for $\left( \sum
A^h_{1/2}(\tau_{\ell^{-}_i})g^h_{\ell^{-}_{ii}}/m_{\ell^{-}_i} \right) \times \left( \sum A^h_{1/2}(\tau_{\ell^{=}_i})g^h_{\ell^{=}_{ii}}/m_{\ell^{=}_i} \right) < 0$, a different sign holds between the $\ell^{-}_i$ and $\ell^{=}_i$ eigenstate contributions to the $h$-diphoton loop. Such a wanted sign configuration can generally be accommodated by controlling the VL Yukawa couplings and masses in the interaction basis\footnote{Including the Yukawa couplings between the SM and the chosen VLLs in the Lagrangian, one could indeed find combinations of four Yukawa couplings whose global sign has a physical impact, since these combinations are invariant under any field redefinition.}. As expected from unitarity and different $\beta$ dependences, it is in contrast a constructive interference which occurs simultaneously in the $H\gamma\gamma$ loops [see eqs.~(\ref{eq:agg})--(\ref{eq:app})], of course for an identical sign configuration of the above product of Yukawa couplings -- now involving $A^H_{1/2}$ factors. The interference sign between up and down VLL contributions to the $A$ loop cannot be deduced generically from the $h,H$ configurations as the involved Yukawa couplings are different (sign difference in fermion interaction basis -- see eq.~\eqref{eq:yukawas}): a VLL model--dependence enters here. Nevertheless, it can be noticed that the up--down VLL interference sign is opposite in the MSSM and considered type II 2HDM for the $A$ loop contribution (sign unchanged for the $H$ loop contribution). Hence, the contributions to the $H$ and $A$ diphoton rates can be enhanced by VLLs with respect to the MSSM, while the $h$ diphoton amplitude can be kept close to its SM value. 

In our 2HDM with the chosen VLL representations, this global situation is guaranteed by the fact that a VLL type (here the $\ell^{=}_i$) couples to the field $\Phi_d^0$ while the other type (the $\ell^{-}_i$) couples to the $\Phi_u^0$ field\footnote{One could e.g.  add  
$\ell^{=}$ leptons coupling as well to $\Phi_u^0$, from additional VLL ${\rm SU(2)_L}$ multiplets. This will however complicate the model without bringing particular interest except more degrees of freedom.}. These two kinds of VLLs will thus couple with relative signs (among each other) being different for the couplings to the light and heavy Higgs eigenstates ($h$, $H$)  and with independently controllable Yukawa couplings ($g_{\ell^{=}_i}$ and $g_{\ell^{-}_i}$). 

In the case of VLQs, this mechanism is expected to be less effective as one cannot simultaneously cancel the contributions to the $h \gamma\gamma$  and $hgg$ loops, since $Q_t\neq Q_b$. To circumvent the problem, a possibility is to consider vector--like multiplets including ${\rm SU(2)_L}$ doublets with higher electric charges, e.g. $(q^{8/3}, q^{5/3})^t_{L/R} ,  q^{8/3}_{L/R} ,  q^{5/3}_{L/R}$ (as realized for example in Ref.~\cite{RS83}). 
Then, gauge invariant Yukawa couplings can be accommodated and the squared charge ratio tends to unity,  $1<(Q_{q^{8/3}} / Q_{q^{5/3}})^2 \simeq 2.5 < (Q_t/Q_b)^2 = 4$, the reason being that the unit charge difference between the two components of a doublet is fixed by the gauge symmetry due to the common hypercharge. An additional advantage of the representations above is to increase even more the VLQ contributions to the $\Phi \gamma \gamma$ loops, due to the individually large charges, $(Q_{q^{5/3}})^2$ and $(Q_{q^{8/3}})^2$~\cite{Bonne}. Note that VLQs can also explain another excess (at the two sigma-level) that has been observed  at the previous LHC run at 8 TeV in associated $t\bar th$ production \cite{HiggsCombo}, as discussed recently~\cite{ttH-VLQ}.

\begin{figure}[!h]
\vspace*{.3cm}
\begin{center}
\resizebox{0.43\textwidth}{!}{\includegraphics{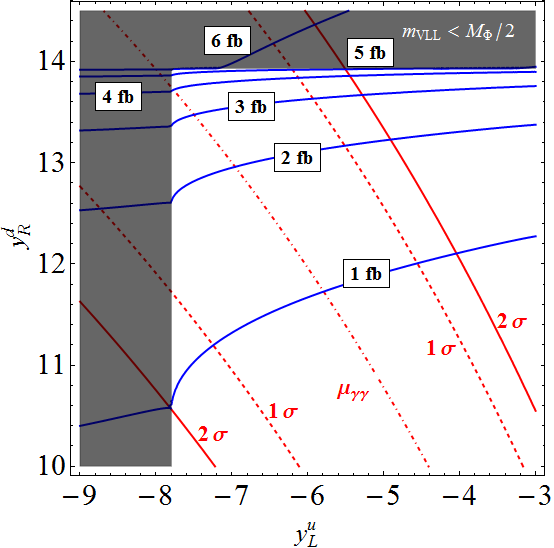}}
\quad
\resizebox{0.43\textwidth}{!}{\includegraphics{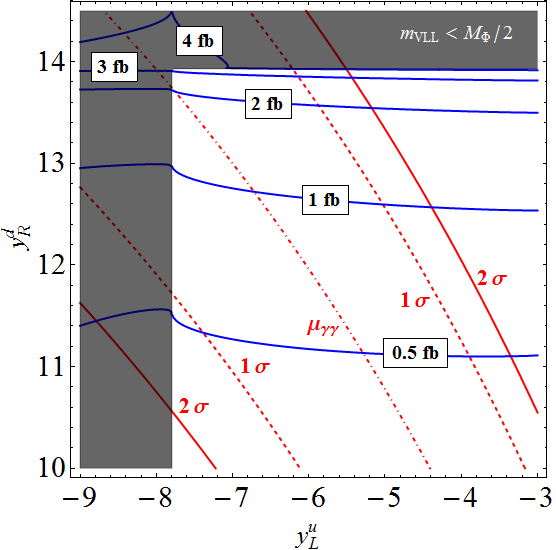}}
\end{center}
\vspace*{-.6cm}
\caption{Contours of constant $\sum \sigma (gg\to \Phi) \times {\rm BR}( \Phi \to \gamma\gamma)$ (blue, in fb) and $\mu_{\gamma\gamma}$ (red) in the $\{y_L^u,y_R^d\}$ plane, for the MSSM case (left) and for the type II 2HDM case (right). The dot-dashed line represents the experimental central value of the $h\to\gamma\gamma$ signal strength, $\mu_{\gamma\gamma} = 1.16 \pm 0.18 \pm 0.15$, while the dashed (solid) lines represent the $1\sigma$ ($2\sigma$) bands. The gray shaded region corresponds to at least one VLL eigenmass being smaller than $\frac12 M_{\Phi} \simeq 375$~GeV. The values of the other parameters are given by $y_R^u= 1.65 $, $y_L^d = 0.4 $, $m_1= 970 $~GeV, $m_2= 330 $~GeV, $m_3= 910 $~GeV.}
\label{Fig:plots}
\vspace*{-3mm}
\end{figure}

We present now some numerical results that can be obtained when considering a specific model.  We consider a particle spectrum with six identical copies\footnote{It is possible that the number of VLL families can be reduced, for same order diphoton rates, if non-vanishing Yukawa couplings between the different families are considered. Note also that diphoton rates around the femtobarn can be reached through the $H$ production with only three VLL generations, in the specific case of negative relative-sign corrections to the SM $h$-diphoton amplitude of $\sim 200\%$ (also compatible with the measured $h$ signal strength).} of the VLL multiplets presented in eq.~\eqref{eq:VLL-ii}. For simplicity, we assume that these six copies do not mix between themselves and that they are described by identical parameter values. In order to maximize the impact of the VLL contributions to the $\Phi \gamma\gamma$ loops, a tempting choice would be to choose the interaction basis parameters in such a way that all the new VLLs have equal masses which are close to the threshold $\frac12 M_{\Phi}$ (i.e. $\tau =1$), where the $A^\Phi_{1/2}$ form factors are close to their maximal value, as shown in Fig.~\ref{Fig:Afactor}. However, such a choice would set to zero
both the up--type (singly charged) and down--type (doubly-charged) VLL contributions to the $A\to\gamma\gamma$ process~\cite{VLQ-india}. Instead, to take at least partly advantage of the sizeable form factor, we arrange the parameter values such that only one  up--like and one down--like VLL per copy have masses close to the threshold. Also, as discussed before, we will only consider the regions of the parameter space where the singly and doubly-charged contributions to the $h\to\gamma\gamma$ loop approximately cancel each other, leaving thus the $h\gamma\gamma$ effective coupling SM-like. This leads automatically to constructive interference in the $H\to\gamma\gamma$ loop.

In Fig.~\ref{Fig:plots}, we present isocontours of $\sum_{\Phi=H,A} \sigma (gg\to \Phi) \times {\rm BR}(\Phi \to \gamma\gamma)$  and the signal strength for the previously observed SM--like state $\mu_{\gamma \gamma}=\sigma (gg\to h)  \times {\rm BR}(h \to \gamma\gamma)/ \sigma (gg\to h)  \times {\rm BR}(h \to \gamma\gamma)|_{\rm SM}$ in the $\{y_L^u,y_R^d\}$ plane, for both the MSSM and type II 2HDM cases. The signal strength for the lighter $h$ boson has been evaluated according to the discussion in Ref.~\cite{ttH-VLQ}: we took the latest combined experimental value obtained at the previous run of the LHC, $\mu_{\gamma\gamma} = 1.16 \pm 0.18$ \cite{HiggsCombo}, and added a theoretical uncertainty of order 15\% \cite{DTHiggs}. We chose to vary the $y_L^u$ and $y_R^d$ parameters because each one is representative for its own sector (up/singly--charged and down/doubly--charged). Also, to avoid a too large width of the scalar resonances, we constrained the VLL eigenmasses to be slightly higher than $\frac12 M_{\Phi} \simeq 375$~GeV. Albeit the high values of the parameters varied in the plot, the mass basis Yukawa couplings have values well below the perturbativity limit of $4\pi$. The Yukawa couplings with the largest values are $g^{A}_{d_{22}} \simeq g^{H}_{d_{22}} \simeq 5.8$. As mentioned earlier, the highest values of the total cross section times branching ratio for the $H$ and $A$ resonances occur in the region where $m_{l^-_1} \simeq m_{l^=_1} \simeq \frac12 M_{\Phi}$. For completeness, we quote the other two VLL masses, which, in the region where $\sigma\times BR$ is maximized, are given by $m_{l^-_2}\simeq 1.3$~TeV and $m_{l^=_2} \simeq 2.1$~TeV. Moreover, in the total diphoton cross section,  $\sum_{\Phi} \sigma (gg\to \Phi) \times {\rm BR}( \Phi \to \gamma\gamma)$, the pseudoscalar contribution is $\sim 5.7 \, (3.6)$ times larger than the $H$ contribution for the MSSM (2HDM) case.

In conclusion, by adding charged VLLs, one can obtain  values of $\sum_{\Phi} \sigma (gg\to \Phi) \times {\rm BR}( \Phi \to \gamma\gamma) \simeq 5$--6~fb (3--4~fb) for the diphoton rate in the MSSM (type II 2HDM) case, while keeping the $h\to\gamma\gamma$ signal strength in agreement with the LHC Higgs data~\cite{HiggsCombo}. By comparing with the value of the diphoton rate for the $H/A$ resonances that can be obtained in our 2HDM scenarii with $\tb =1$,  i.e. $
\sigma \times {\rm BR} \approx 1.5\times 10^{-2}$~fb, we see that the VLL loop contributions allow to enhance the decay rate of the $A$ and $H$ bosons to $\gamma\gamma$ final states by a factor of $\mathcal{O}(100)$. Such an important enhancement is due to $(i)$ the high electric charges of the VLLs, $(ii)$ the several VLL families and $(iii)$ the fact that half of the VLLs have masses $ M \gtrsim\frac12 M_{\Phi}$, for which  the form factors attain their maximal values.


\subsection*{4. Conclusions}

The first searches performed at the new LHC with a center of mass energy of 13 TeV, albeit with a moderate accumulated luminosity, look very promising as the ATLAS and CMS collaborations have reported the observation of a diphoton resonance at an invariant mass of about $750$~GeV. The significance of the signal is well below the required five standard deviations and it can well be a statistical fluctuation. It is nevertheless tempting to consider the possibility that it is the first sign of new physics beyond the SM. 

We have investigated the possibility that the diphoton resonance is one of the heavy neutral CP--even or CP--odd Higgs particles (and in fact a superposition of the two) that arise in two--Higgs doublet scenarios that are considered as a straightforward extension of the SM and widely studied, especially in the context of supersymmetric theories like the MSSM. However, such a strong diphoton signal cannot be achieved in the usual versions of 2HDMs and the MSSM and additional charged particles should contribute to the loop induced production and decay processes. 

We have thus considered the possibility that these new particles are vector--like quarks and leptons that couple strongly to the heavy Higgs bosons. We have shown that, for instance, six families of VL leptons can easily enhance the diphoton rate of the 750 GeV resonance to accommodate the observed signal, without affecting the properties of the light 125 Higgs boson, whose couplings have been measured to be SM--like.

If this diphoton excess is not a statistical fluctuation and is indeed confirmed by future data as being a real physics signal, it will have far reaching consequences. The new physics to which the signal is connected must be extremely rich since, at the same time,  it implies the existence of a heavy Higgs--like resonance which is already a very interesting signal for physics beyond the SM and, very likely  also, of some other new particle content as to increase the production rate of the 750 GeV resonance in the diphoton channel. These new particles can be light enough to be directly produced and studied in detail at the LHC with significantly 
higher luminosities than those collected so far. This would open a very rich and 
exciting era for high energy particle physics.\bigskip   

\noindent {\bf Acknowledgements:} Discussions with Luciano Maiani and Fran\c{c}ois Richard
and correspondence with Xiao-Fang Han are gratefully acknowledged. A.D. thanks the CERN Theory Unit for its hospitality. This work is supported by the ERC advanced grant Higgs@LHC. G.M. is supported by the ``Institut Universitaire de France'' and the European Union FP7 ITN Invisibles.


\baselineskip=14pt
\begin{thebibliography}{999}

\bibitem{annonce} Jim Olsen for CMS and Marumi Kado for ATLAS, talks given 
at ``ATLAS and CMS physics results from Run 2", CERN, 15/12/2005.  

\bibitem{Landau-Yang} L. Landau,  Dokl. Akad. Nauk Ser. Fiz. 60 (1948) 207;
C. Yang, Phys. Rev. 77 (1950) 242. 

\bibitem{Latest-13TeV} ATLAS collaboration (G.~Aad  et al), arXiv:1512.01530 [hep-ex]; 
CMS collaboration (V.~Khachatryan et al.), arXiv:1512.01224 [hep-ex]. 

\bibitem{LHC-XS} See the LHC Higgs XS-WG page, {\tt twiki.cern.ch/twiki/bin/view/LHCPhysics/LHCHXSWG}; see also, G. Salam and A. Weiler, {\tt cern.ch/collider-reach/}. 

\bibitem{ATLAS-earlier}  ATLAS Collaboration (G.~Aad et al.),  Phys. Rev. Lett.  113 (2014)  no. 17, 171801. 

\bibitem{CMS-earlier} CMS Collaboration (V.~Khachatryan et al.), Phys. Lett. B750 (2015)  494.  

\bibitem{2HDM}  For a review on 2HDMs, see G. Branco et al.,  Phys. Rept. 516 (2012) 1. 

\bibitem{HHG} \mbox{J. Gunion, H. Haber, G. Kane and S. Dawson, ``The Higgs Hunter's
Guide", Reading 1990.} 

\bibitem{Review2} A.~Djouadi,  Phys. Rept. 459 (2008) 1. 

\bibitem{VLQs} For recent reviews see: S. Ellis, R. Godbole, S. Gopalakrishna and J. Wells, JHEP 1409 (2014) 130; J.A.~Aguilar-Saavedra et al., Phys. Rev. D88 (2013)  094010; 
A. Azatov et al., Phys. Rev. D85 (2012) 115022. 

\bibitem{HiggsCombo} ATLAS and CMS Collaborations, ATLAS-CONF-2015-044.

\bibitem{alignment} See e.g. M. Carena et al., JHEP 1404 (2014) 015. 

\bibitem{hMSSM} A. Djouadi et al., Eur. Phys. J. C73 (2013) 2650;
JHEP 1506 (2015) 168.  

\bibitem{LHC-tautau} ATLAS collaboration (G.~Aad  et al), JHEP 1411 (2014) 056;  CMS collaboration (V.~Khachatryan et al.), JHEP 1410 (2014) 160.  

\bibitem{Review1} A.~Djouadi,  Phys. Rept. 457 (2008) 1. 

\bibitem{SDGZ} M. Spira, A. Djouadi, D. Graudenz and P. Zerwas, Nucl. Phys. B453 (1995) 17.

\bibitem{hdecay}  The numerical analyses on the Higgs decays are performed using the program {\tt HDECAY}: A.~Djouadi, J.~Kalinowski and M.~Spira, Comput. Phys.
Commun. 108 (1998) 56; A. Djouadi, M. Muhlleitner and M. Spira, Acta. Phys. Polon. 
B38 (2007) 635. 

\bibitem{LHC-ttbar} ATLAS Collaboration (G.~Aad  et al), JHEP 1508 (2015) 148; 
CMS Collaboration (V.~Khachatryan  et al.), arXiv:1506.03062 [hep-ex].

\bibitem{PDG} Particle Data Group (K. Olive et al.), Chin. Phys. C38 (2014) 090001.

\bibitem{golden} For a recent account: A. Djouadi, J. Quevillon and
R. Vega-Morales, arXiv:1509.03913 [hep-ph].  

\bibitem{DCHiggs} See e.g. A. Arhrib et al.,  JHEP 1204 (2012) 136; 
A. Akeroyd and S. Moretti, Phys. Rev. D86 (2012) 035015.  

\bibitem{fourth} A. Djouadi and A. Lenz, Phys. Lett. B715 (2012) 310;
A. Denner et al., Eur.Phys.J. C72 (2012) 1992;
E. Kuflik, Y. Nir and T. Volansky, Phys. Rev. Lett. 110 (2013) 091801.  

\bibitem{SUSY+VLQ} Z.~Lalak, M.~Lewicki and J.D.~Wells, Phys.\ Rev.\ D 91 (2015) 9,  095022. 

\bibitem{VLLEP}  See e.g. 
K. Nakamura et al. (Particle Data Group), J. Phys. G37 (2010)  075021; 
A. Joglekar, P. Schwaller and C. E. Wagner, JHEP 1212 (2012) 064; ATLAS Collaboration (G.~Aad  et al.),  JHEP  1509 (2015) 108.

\bibitem{LHC-VLQS} ATLAS collaboration (G.~Aad  et al.), JHEP 08 (2015) 105;
CMS collaboration (V.~Khachatryan  et al.), Phys. Lett. B729 (2014) 149.

\bibitem{VLQ-india} See e.g. S. Gopalakrishna,  S. Mitra and G. Moreau, JHEP 1408 (2014) 079; 
S.~Gopalakrishna, T.~S.~Mukherjee and S.~Sadhukhan, arXiv:1504.01074 [hep-ph].

\bibitem{DM} The possibility that one of the heavy neutrinos is dark matter
will be studied in a separate paper: G. Arcadi et al., to appear.

\bibitem{RS83} C. Bouchart and G. Moreau, Nucl. Phys. B810 (2009) 66.  

\bibitem{Bonne} N. Bonne and G. Moreau, Phys. Lett. B717 (2012) 409; G. Moreau
Phys. Rev. D87 (2013) 015027.  

\bibitem{ttH-VLQ} A.~Angelescu, A.~Djouadi and G.~Moreau, arXiv:1510.07527 [hep-ph]. 

\bibitem{DTHiggs} J. Baglio and A. Djouadi, JHEP 1103 (2011)  055; 
A. Djouadi and G. Moreau, Eur. Phys. J. C73  (2013) 2512;
S.~Fichet and G.~Moreau, arXiv:1509.00472 [hep-ph]. 


\end{thebibliography}
\end{document}